\documentclass{article}
\usepackage[onecolumn]{emulateapj}
\usepackage{graphicx}
\usepackage{timesfonts} 

\submitted{
\sc The Astrophysical Journal, \rm 528:\#\#\# -- \#\#\#, 2000, January 1 
   (tentatively)}

\makeatletter
\@ifundefined{chapter}{\def\thebibliography#1{
\section*{References}
  \list
  {\relax}{\setlength{\labelsep}{0em}
        \setlength{\itemindent}{-\bibhang}
        \setlength{\itemsep}{\parskip}
        \setlength{\parsep}{0pt}
        \setlength{\leftmargin}{\bibhang}}
    \def\newblock{\hskip .11em plus .33em minus .07em}
    \sloppy\clubpenalty4000\widowpenalty4000
    \sfcode`\.=1000\relax}}%
{\def\thebibliography#1{
  \list
  {\relax}{\setlength{\labelsep}{0em}
        \setlength{\itemindent}{-\bibhang}
        \setlength{\itemsep}{\parskip}
        \setlength{\parsep}{0pt}
        \setlength{\leftmargin}{\bibhang}}
    \def\newblock{\hskip .11em plus .33em minus .07em}
    \sloppy\clubpenalty4000\widowpenalty4000
    \sfcode`\.=1000\relax}}

\newlength{\bibhang}
\makeatother

\def\title#1{{\thispagestyle{apjbot}
  \subtitle
  \vspace{12mm} 
  \center
\begin{minipage}{\abstrwidth}
\center\uppercase{#1}\endcenter
\end{minipage}\endcenter}}

\newcommand{\gray}{$\gamma$-ray\ }
\newcommand{\grays}{$\gamma$-rays\ }

\def\fwa{80mm}   
\def\fwb{80mm}   
\def\fwc{68mm}   
\def\fwd{80mm}   
\def\hs{15mm}
\renewcommand{\baselinestretch}{0.95}

\lefthead{MOSKALENKO AND STRONG}
\righthead{ANISOTROPIC INVERSE COMPTON SCATTERING IN THE GALAXY}

\begin{document}

\title{Anisotropic inverse Compton scattering in the Galaxy}

\author{Igor V.~Moskalenko\altaffilmark{1,2} and 
        Andrew W.~Strong\altaffilmark{1}}

\affil{\altaffilmark{1}Max-Planck-Institut f\"ur extraterrestrische Physik,
   Postfach 1603, D-85740 Garching, Germany}
\affil{\altaffilmark{2}Institute for Nuclear Physics, 
   M.V.Lomonosov Moscow State University, 119 899 Moscow, Russia}



\begin{abstract}

The inverse Compton scattering of interstellar photons off cosmic-ray
electrons seems to play a more important r\^ole in the generation of
diffuse emission from the Galaxy than thought before.  The background
radiation field of the Galaxy is highly anisotropic since it is
dominated by the radiation from the Galactic plane.  An observer in the
Galactic plane thus sees mostly head-on scatterings even if the
distribution of the cosmic-ray electrons is isotropic.  This is
especially evident when considering inverse Compton scattering by
electrons in the halo, i.e.\ the diffuse emission at high Galactic
latitudes.  

We derive formulas for this process and show that the anisotropy of the
interstellar radiation field has a significant effect on the intensity
and angular distribution of the Galactic diffuse $\gamma$-rays, which
can increase the high-latitude Galactic \gray flux up to
40\%.  This effect should be taken into account when calculating the
Galactic emission for extragalactic background estimates.

\end{abstract}

\vskip 1ex

\section{Introduction}
Recent studies show that inverse Compton scattering (ICS) of
interstellar photons off cosmic-ray (CR) electrons seems to play a more
important r\^ole in generation of diffuse \gray emission from the
Galaxy than thought before (e.g., \cite{PorterProtheroe97},
\cite{PohlEsposito98}, \cite{MS98b}).  The spectrum of Galactic \grays
as measured by EGRET shows enhanced emission above 1 GeV in comparison
with calculations based on locally measured proton and electron spectra
assuming the same spectral shape over the whole Galaxy
(\cite{Hunter97}, \cite{Gralewicz97}, \cite{Mori97}, \cite{SM97},
\cite{MS98a}).  The \gray observations therefore indicate that their
spectra on the large scale in the Galaxy could be different.  Harder CR
spectra could provide better agreement, but the \gray data alone cannot
discriminate between the $\pi^0$-decay and inverse Compton
explanations. The hard electron spectrum hypothesis seems to be more
likely due to the probably clumpy distribution of electrons at high
energies (e.g., \cite{PohlEsposito98}), while the hard nucleon spectrum
can be excluded at the few $\sigma$ level on the basis of antiproton
data above 3 GeV and positron measurements (\cite{MS98b},
\cite{MSR98}).  If the interstellar electron spectrum is indeed harder
than that measured directly, then the ICS is a main contributor to
the diffuse Galactic emission above several MeV and can account for the
famous `GeV excess'.

The origin and spectrum of the Galactic diffuse \gray emission has a
strong impact also on the extragalactic studies and cosmological
implications.  The origin of the truly extragalactic \gray background
is still unknown.  The models discussed range from the primordial black
hole evaporation (e.g., \cite{PageHawking76}) and annihilation of
exotic particles in the early Universe (e.g., \cite{ClineGao92}) to the
contribution of unresolved discrete sources such as active galaxies
(e.g., \cite{Sreekumar98}), while the spectrum of the extragalactic
emission itself is the main uncertain component of these models.  The
latter can be addressed only by the accurate study of the Galactic
diffuse emission.  Additionally, it is relevant to dark matter studies
such as the search for signatures of WIMP annihilation, which is also a
potential source of $\gamma$-rays, positrons, and antiprotons in the
Galactic halo (see \cite{Jungman96} for a review).

The calculations of the ICS in most studies are based on the formula
obtained by Jones (1968) which remains a good approximation when
considering isotropic scattering.  However, the interstellar radiation
field (ISRF) (apart from the cosmic microwave background) is
essentially anisotropic.  The stellar and dust emission originates in
the Galactic plane and preferentially scatters back to the observer
also in the plane.  Therefore, the observer sees mostly head on
collisions even if the electron distribution is isotropic.  Though
qualitatively such an effect was pointed out many years ago (e.g.,
\cite{WorrallStrong77}), the magnitude of this effect has not been
evaluated accurately in the literature so far.  Since ICS is one of the
most important components of the diffuse Galactic \gray emission its
accurate calculation is of interest for astrophysical studies.

We have developed a propagation code (`GALPROP') which aims to
reproduce self-consistently observational data of many kinds related to
CR origin and propagation: direct measurements of nuclei, antiprotons,
electrons and positrons, $\gamma$-rays, and synchrotron radiation
(\cite{SM98}, \cite{MS98a}, \cite{MSR98}).  In this paper we evaluate
the effect of anisotropy in the ICS of CR electrons off interstellar
photons and present corresponding formulas, which serve a basis for our
detailed analysis of the Galactic diffuse emission in an accompanying
paper (\cite{SMR99}).  We make calculations with a realistic ISRF and
also a simplified distribution of photons (thin emitting disk), which
allows for simpler analytical treatment.  The latter shows the effect
without the uncertainty introduced by calculation of the real ISRF
(which is itself a complicated task).  For interested users our model
including software and result datasets is available in the public
domain on the World Wide
Web\footnote{http://www.gamma.mpe--garching.mpg.de/$\sim$aws/aws.html}.

Throughout the paper the units $\hbar=c=m_e=1$ are used.

\section{Inverse Compton scattering in anisotropic photon field \label{IC}}
We start from the general formula for the rate of photon-particle 
interactions (\cite{Weaver76}):
\begin{equation}
\label{IC.1}
R = 
   n_e n_\gamma \int d\vec{p}_\gamma \int d\vec{p}_e\, f_e(\vec{p}_e) 
   f_\gamma(\vec{p}_\gamma) \frac{p'_\gamma}{\gamma p_\gamma}
   \sigma(p'_\gamma),
\end{equation}
where $n_e, n_\gamma$ are the electron and photon number densities,
respectively, $\vec{p}_e, \vec{p}_\gamma$ are the momenta,
$f_e(\vec{p}_e), f_\gamma(\vec{p}_\gamma)$ are the corresponding
distribution functions in the laboratory system (LS) (the normalization
being chosen so that $\int f_{e,\gamma} (\vec{p}_{e,\gamma})\,
d\vec{p}_{e,\gamma} = 1$), $\gamma$ is the electron Lorentz factor,
$\sigma$ is the cross section, and prime marks the electron-rest-system
(ERS) variables.
In the following treatment the electron distribution is assumed
isotropic.

\subsection{Scattering off a single electron}
For an electron energetic enough the incoming photons in the ERS are
seen as a narrow beam, $\sim 1/\gamma$ wide, in the forward direction.
Therefore, we adopt Jones' (1968) approximation considering the
incoming photons as a unidirectional beam in the ERS.  Relaxing this
assumption leads to overcomplexity of the final formulas and their
derivation becomes lengthy and laborious while has little effect on the
final result.  An equivalent approach is valid for the LS, the photons
are mostly scattered forward in a cone of $\sim 1/\gamma$ wide and can
be also considered as unidirectional.  The latter will be used in
Section \ref{kinematics} to set up kinematic limits on the angles
involved.

\begin{figure}[t]
\centerline{
      \hspace{5mm}
      \includegraphics[width=\fwc]{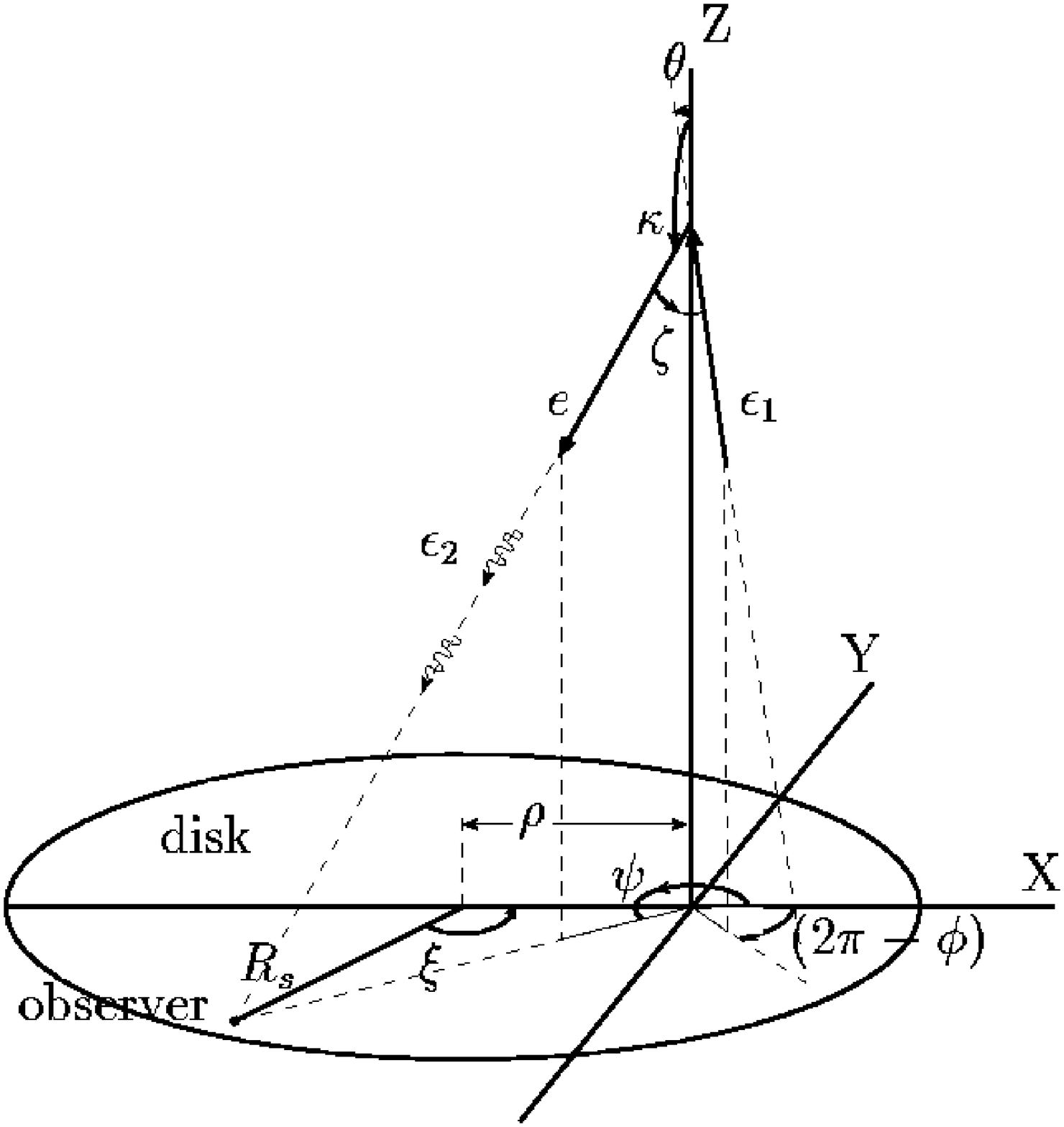}
      \hspace{\hs}
      \includegraphics[width=\fwa]{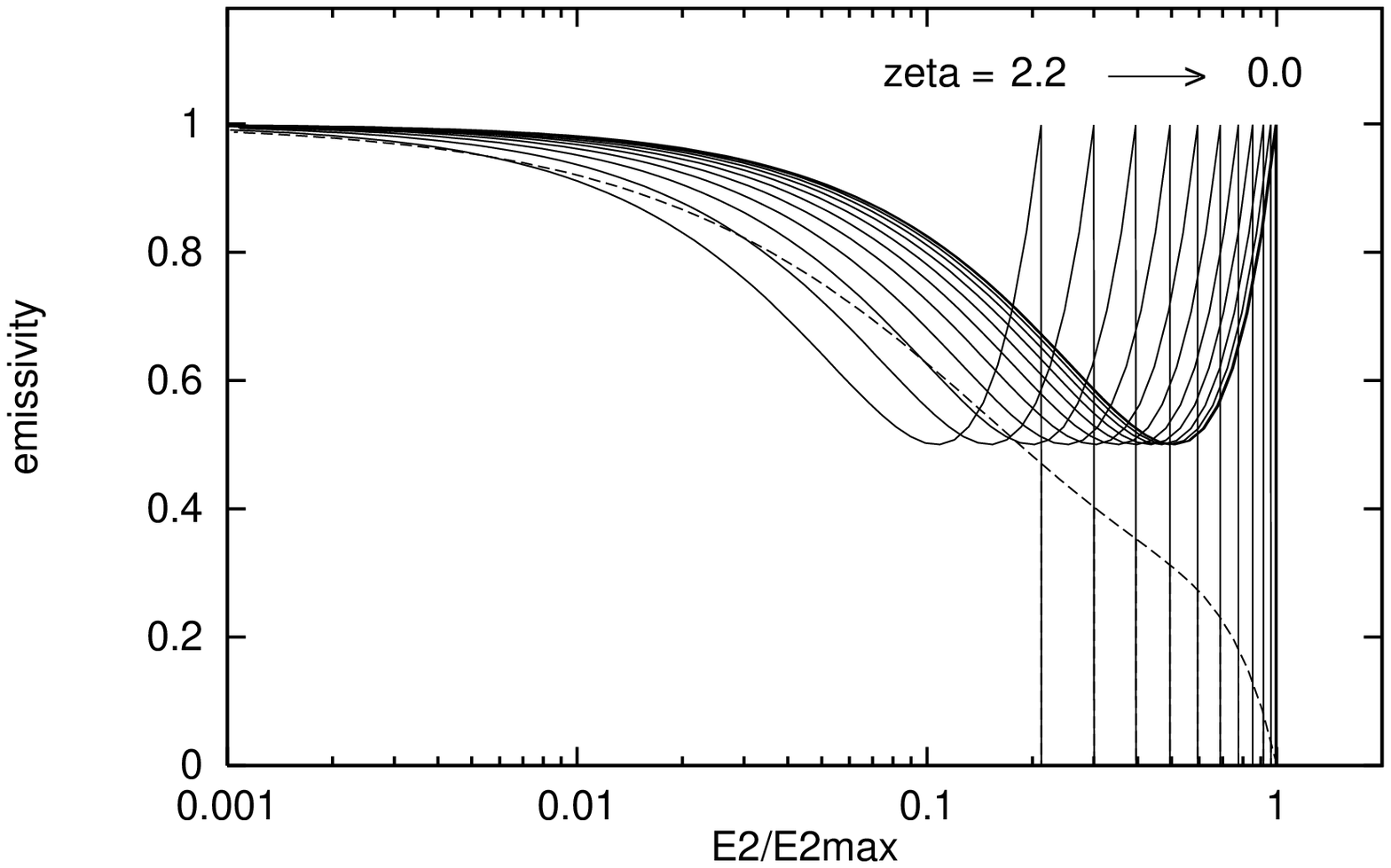}
}
\parbox{89mm}{%
\figcaption[fig1.ps]{ 
Angles involved in the ICS process: $(\theta,\phi)$, polar and
azimuthal angles of the incoming photon, $(\kappa,\psi)$, the same
for the electron, $R_s$, observer's position. See text
for more details.
\vspace{2\baselineskip}\vspace{1pt}
\label{fig1}}
}\hspace{7mm}
\parbox{89mm}{%
\figcaption[fig2.ps]{
The spectra of upscattered photons for isotropic scattering
(eq.~[\ref{IC.32}]: dashes) and for the case of a parallel beam of
target photons (anisotropic scattering, eq.~[\ref{IC.8}]: solid lines)
for $\gamma = 10^5$ and $\epsilon_1 = 10^{-7}$.  In the latter case
spectra are shown for several angles between the source and the
observer, $\zeta= 0\, (0.2)\, 2.2$ radians.  The emissivities shown are
divided by $0.002 \pi r_e^2$.
\label{fig2}}
}
\end{figure}

First consider monoenergetic electrons with distribution
\begin{equation}
\label{IC.2}
f_e(\vec{p}_e) = 
   \frac{1}{4\pi p^2_e} \delta(p_e-p),
\end{equation}
where $\delta(x)$ is the Dirac delta function.  In terms of the
electron Lorentz factor this gives ($\Omega_e$ is the solid angle)
\begin{equation}
\label{IC.3}
f_e(\gamma_1, \Omega_e) = 
   \frac{1}{4\pi \gamma_1^2} \delta(\gamma_1-\gamma),
\end{equation}
where $p=\beta\gamma$, and $\beta = 1$ is the electron speed.
Consider also monoenergetic target photons with distribution given by
\begin{equation}
\label{IC.4}
f_\gamma(\epsilon_\gamma, \Omega_\gamma) = 
   Q_\gamma(\theta,\phi)
   \frac{1}{\epsilon^2_\gamma} \delta(\epsilon_\gamma-\epsilon_1),
\end{equation}
where $\epsilon_\gamma= p_\gamma$ is the photon energy,
$Q_\gamma(\theta,\phi)$ is the angular distribution of photons ($\int
d\Omega_\gamma\, Q_\gamma = 1$) at a particular spatial point, and
$(\theta, \phi)$ are the photon polar angle and the azimuthal angle,
respectively (Fig.~\ref{fig1}).

The Klein-Nishina cross section is given by (\cite{JauchRohrlich76})
\begin{equation}
\label{IC.5}
\frac{d\sigma}{d\epsilon'_2 d\cos\eta'} = \pi r_e^2
   \left(\frac{\epsilon'_2}{\epsilon'_1}\right)^2
   \left(\frac{\epsilon'_2}{\epsilon'_1}+\frac{\epsilon'_1}{\epsilon'_2}
   -\sin^2\eta'\right)
   \delta\left(\epsilon'_2-\frac{\epsilon'_1}{1+\epsilon'_1(1-\cos\eta')}
   \right),
\end{equation}
where $r_e$ is the classical electron radius,
$\epsilon'_1, \epsilon'_2$ are the ERS energies of the incoming
and scattered photons, respectively, and $\eta'$ is the scattering 
angle in the ERS.

The upscattered photon distribution over the LS energy, $\epsilon_2$,
as obtained from eq.~(\ref{IC.1}) is
\begin{equation}
\label{IC.6}
\frac{dR}{d\epsilon_2} = 
   \int d\cos\eta'
   \int d\epsilon_\gamma d\Omega_\gamma 
   \int d\gamma_1 d\Omega_e\, 
   f_e(\gamma_1,\Omega_e) f_\gamma(\epsilon_\gamma,\Omega_\gamma) 
   \epsilon_\gamma^2 \gamma_1^2
   \frac{\epsilon'_\gamma}{\gamma_1 \epsilon_\gamma}\,
   J\!\left(\frac{\epsilon'_2}{\epsilon_2}\right)\,
   \frac{d\sigma}{d\epsilon'_2 d\cos\eta'},
\end{equation}
where $J(\epsilon'_2/\epsilon_2)=\epsilon'_2/\epsilon_2$ is the
Jacobian (ERS $\to$ LS), $\epsilon'_2$ is a function of $\cos\eta'$ and
LS variables, $p'_\gamma= \epsilon'_\gamma$, and we have put
$n_{e,\gamma}=1$.

Taking into account the $\delta$-functions in
eqs.~(\ref{IC.3}),(\ref{IC.4}), the integration over $d\epsilon_\gamma,
d\gamma_1, d\Omega_e$, is trivial, and yields
\begin{equation}
\label{IC.7}
\frac{dR(\gamma,\epsilon_1)}{d\epsilon_2} = 
   \frac{\pi r_e^2}{\gamma\epsilon_1\epsilon_2} 
   \int d\cos\eta' \int_{\Omega_\gamma} d\Omega_\gamma\,
    Q_\gamma(\Omega_\gamma)
   \frac{{\epsilon'_2}^3}{\epsilon'_1}\,
   \left(\frac{\epsilon'_2}{\epsilon'_1}+\frac{\epsilon'_1}{\epsilon'_2}
   -\sin^2\eta'\right)
   \delta\left(\epsilon'_2-\frac{\epsilon'_1}{1+\epsilon'_1(1-\cos\eta')}
   \right),
\end{equation}
where $\epsilon'_2 = \epsilon_2/[\gamma(1-\beta\cos\eta')]$, and
$\eta'=\pi$ for forward scattering.

The integration over $d\cos\eta'$ is also trivial ($\gamma> \epsilon_2$)
\begin{equation}
\label{IC.8}
\frac{dR(\gamma,\epsilon_1)}{d\epsilon_2} = 
   \frac{\pi r_e^2}{\epsilon_1(\gamma-\epsilon_2)^2}
   \int_{\Omega_\gamma} d\Omega_\gamma\, Q_\gamma(\Omega_\gamma)
   \left[2 -2\frac{\epsilon_2}{\gamma}\left(\frac{1}{\epsilon'_1}+2\right)
   +\frac{\epsilon_2^2}{\gamma^2}\left(\frac{1}{{\epsilon'_1}^2}
      +2\frac{1}{\epsilon'_1}+3\right)
   -\frac{\epsilon_2^3}{\gamma^3}\right],
\end{equation}
where 
\begin{equation}
\label{IC.36}
\epsilon_2 \le 2\gamma \epsilon'_1/ (1+2\epsilon'_1), \qquad
\epsilon'_1 = \epsilon_1\gamma(1+\beta\cos\zeta),
\end{equation}
$\zeta$ is the LS angle between the momenta of the electron and
incoming photon ($\zeta=0$ for head-on collisions, see
Fig.~\ref{fig1}), and
\begin{equation}
\label{IC.41}
\epsilon_{2\max}= 4\epsilon_1\gamma^2/(1+4\epsilon_1\gamma)
\end{equation}
is the maximal energy of the upscattered photons.

Eq.~(\ref{IC.8}) gives the spectrum of the upscattered photons per
single electron and single target photon, and is the basic formula in
our derivation.  For an isotropic distribution of photons this can be
compared with the classical Jones' (1968) result.  Let the $z$-axis be
\linebreak[3] 
(anti-) parallel to the electron momentum so that $\zeta
\equiv \theta$ and $d\Omega_\gamma= 2\pi d\cos\zeta$; $Q_\gamma =
1/4\pi$ for the isotropic photon distribution.  Changing the
integration variable from $\zeta$ to $\epsilon'_1$ gives
\begin{equation}
\label{IC.35}
\frac{dR(\gamma,\epsilon_1)}{d\epsilon_2} = 
   \frac{\pi r_e^2}{2\epsilon_1^2\gamma(\gamma-\epsilon_2)^2}
   \int^{2\gamma\epsilon_1}_{\epsilon_2/2(\gamma-\epsilon_2)} d\epsilon'_1\, 
   \left[2 -2\frac{\epsilon_2}{\gamma}\left(\frac{1}{\epsilon'_1}+2\right)
   +\frac{\epsilon_2^2}{\gamma^2}\left(\frac{1}{{\epsilon'_1}^2}
      +2\frac{1}{\epsilon'_1}+3\right)
   -\frac{\epsilon_2^3}{\gamma^3}\right].
\end{equation}
The result, after some rearrangement, exactly coincides with the
well-known formula (\cite{Jones68})
\begin{equation}
\label{IC.32}
\frac{dR_{iso}}{d\epsilon_2}(\gamma,\epsilon_1) = 
   \frac{2\pi r_e^2}{\epsilon_1\gamma^2}
   \left[2 q'\ln q' +(1+2q')(1-q')+\frac{1}{2}
   \frac{(4\epsilon_1\gamma q')^2}{(1+4\epsilon_1\gamma q')}(1-q')\right],
\end{equation}
where $q'=\epsilon_2/[4\epsilon_1\gamma^2(1-\epsilon_2/\gamma)]$ and
$1/4\gamma^2<q'\le1$.

\placefigure{fig1}
\placefigure{fig2}

Fig.~\ref{fig2} compares the spectra of upscattered photons from a
parallel photon beam (using eq.~[\ref{IC.8}]) with the isotropic case
(eq.~[\ref{IC.32}]).  The spectra are shown versus
$\epsilon_2/\epsilon_{2\max}$ for several angles $\zeta$ (in the
anisotropic case).  The other parameters are:  $\gamma = 10^5$ and
$\epsilon_1 = 10^{-7}$.  The spectra are different in different
directions, but all are peaked at the energy which corresponds to
highest possible energy of the upscattered photon.  The difference
between the two cases is maximal when $\zeta\approx 0$.

The interpretation of the effect is straightforward and becomes obvious
for the geometry with the source and the observer located at the
same point ($\zeta= 0$).  In this position the observer sees mostly
those photons which suffered head on collision and have been scattered
back, and thus have the highest possible energy $\epsilon_2 =
\epsilon_{2\max}$.  In the case of `isotropic' scattering the number of
such photons is negligible.  Correctly averaged over the viewing angles
eq.~(\ref{IC.8}) will give again the same result as eq.~(\ref{IC.32}).

\subsection{Integration over the solid angle $\Omega_\gamma$}
In principle, eqs.~(\ref{IC.1}),(\ref{IC.8}), and (\ref{IC.36}) are
sufficient to compute the anisotropic ICS for any photon field.
However, the computation can be facilitated using analytical limits on
the angles over which to integrate.

We distinguish kinematic and geometrical restrictions on the angles
involved. These are two independent conditions which should be
satisfied simultaneously.  The kinematics gives the range of angles
$\zeta\le \zeta_0$ for which a photon of energy $\epsilon_1$ can be
upscattered into $\epsilon_2$.  For the case of a disk source the
geometry defines a range of angles so that only the disk photons are
considered. For both conditions we derive analytical formulas for the
integration limits.

\subsubsection{Kinematic limits on $\zeta$ \label{kinematics}}
Using eq.~(\ref{IC.36}) gives the lower limit on $\cos\zeta$
\begin{equation}
\label{IC.12}
\cos\zeta \ge \cos\zeta_0 \equiv
   \frac{\epsilon_2}{2\epsilon_1\gamma(\gamma-\epsilon_2)}-1.
\end{equation}
In terms of polar and azimuthal angles of the electron $(\kappa,
\psi)$ and incoming photon $(\theta, \phi)$ (see Fig.~\ref{fig1})
\begin{equation}
\label{IC.13}
\cos\zeta = -\cos\kappa\cos\theta+\sin\kappa\sin\theta\cos(\phi-\psi).
\end{equation}
Eqs.~(\ref{IC.12}) and (\ref{IC.13}) place restrictions on possible
combinations of the photon angles $(\theta, \phi)$:
\begin{eqnarray}
\label{IC.14}
&\pi-\kappa-\zeta_0 \le  \theta  \le \pi-\kappa+\zeta_0,&\nonumber \\
&\psi-\arccos\Upsilon \le  \phi  \le \psi+\arccos\Upsilon,&
\end{eqnarray}
where 
\begin{equation}
\label{IC.15}
\Upsilon=\frac{\cos\zeta_0+\cos\kappa\cos\theta}
   {\sin\kappa\sin\theta}.
\end{equation}
In the case $|\Upsilon| \ge 1$,
\begin{eqnarray}
\label{IC.16}
\begin{array}{cl}
\psi-\pi \le \phi \le \psi+\pi, &\mbox{for } \Upsilon \le -1;\\
\phi \in \emptyset,             &\mbox{for } \Upsilon \ge +1.
\end{array}
\end{eqnarray}

\subsubsection{Geometrical limits for a disk source}

In the case of a thin disk we can derive analytically the integration
limits (eq.~[\ref{IC.8}]), which come from the position of the electron
above the disk plane.  We consider a coordinate frame centered on the
electron projection on the $(x,y)$-plane of the disk and with $z$ the
altitude of the electron above the disk (Fig.~\ref{fig1}). Then
the coordinates of any point $(x,y)$ in the disk plane can be expressed
in terms of angles ($\theta, \phi$): $x=z\tan\theta\cos\phi$ and
$y=z\tan\theta\sin\phi$, where the angles $\theta$ and $\phi$ are
essentially the same as in eqs.~(\ref{IC.4})--(\ref{IC.8}).
From the disk equation $(x+\rho)^2+y^2\le R_d^2$ one can derive
\begin{equation}
\label{IC.17}
(z\tan\theta\sin\phi)^2 \le R_d^2-(z\tan\theta\cos\phi+\rho)^2,
\end{equation}
where $(x=-\rho,y=0,z=0)$ are the coordinates of the disk center ($\rho
\ge 0$), and $R_d$ is the disk radius. From this it immediately follows
that
\begin{equation}
\label{IC.18}
\begin{array}{rcl}
& c_1 \le \cos\theta \le c_2; & \\
& \phi_1 \le \phi \le \phi_2, &
\end{array}
\end{equation}
where 
\begin{equation}
\label{IC.19}
\begin{array}{ll}
c_1 = z/\sqrt{z^2+(\rho + R_d)^2}; &
c_2 =
   \left\{
   \begin{array}{ll}
   z/\sqrt{z^2+(\rho - R_d)^2}, & \mbox{for } \rho  > R_d;\\
   1,                           & \mbox{for } \rho\le R_d;
   \end{array}
   \right. \\
\phi_1 = \arccos\tau; &
\phi_2 = 2\pi-\arccos\tau,
\end{array}
\end{equation}
and 
\begin{equation}
\label{IC.20}
\tau = (R_d^2-\rho^2-z^2\tan^2\theta)/(2 z\rho\tan\theta).
\end{equation}
In the case $|\tau| \ge 1$,
\begin{eqnarray}
\label{IC.40}
\begin{array}{cl}
0 \le \phi \le 2\pi, &\mbox{for } \tau \ge +1;\\
\phi \in \emptyset,  &\mbox{for } \tau \le -1.
\end{array}
\end{eqnarray}

The actual integration solid angle in eq.~(\ref{IC.8}) is, therefore,
the intersection of the solid angles given by both conditions,
eqs.~(\ref{IC.14})--(\ref{IC.16}) and
eqs.~(\ref{IC.18})--(\ref{IC.40}).

For convenience,
the angles $\kappa$ and $\psi$ can be expressed in terms of the
distance of the observer from the center of the disk, $R_s$, and the angle
$\xi$ between $R_s$ and $\rho$ relative to the disk center:
\begin{eqnarray}
\label{IC.21}
&&\kappa = \pi-\arctan(a/z),\nonumber\\
&&\psi   = \left\{
\begin{array}{lll}
\displaystyle{\pi+\arccos\left(\frac{\rho-R_s\cos\xi}{a}\right)}, &
   \mbox{for} & \ \ \ 0\le\xi\le\pi,\\
\displaystyle{\pi-\arccos\left(\frac{\rho-R_s\cos\xi}{a}\right)}, &
   \mbox{for} & -\pi\le\xi\le 0,
\end{array}
\right.
\end{eqnarray}
where $a = \sqrt{(R_s-\rho\cos\xi)^2+(\rho\sin\xi)^2}$.

\subsubsection{Angular distribution of the background photons}

The angular distribution of the background photons depends essentially
on the disk properties. We distinguish three extreme cases: transparent
disk (a collection of points sources, each of which emits
isotropically: Galactic plane), emitting surface (which emits $\propto
\cos\theta$: stellar surface, accretion disk), and a hypothetical
intermediate case of an `isotropic' disk.  (The hypothetical
`isotropic' disk is considered here, since it provides an isotropic,
within the solid angle $\Omega_\gamma$, distribution of background
photons.)  One can easily show that in these cases the angular
distribution of photons at some particular point\footnote{Note, we
consider the coordinate frame where the $(x,y)$-plane coincides with
the disk plane and the $z$-axis passes through the position of the
electron, not the disk center!} $(x=0,y=0,z)$ above the disk can be
expressed as follows
\begin{equation}
\label{IC.22}
Q_\gamma(\Omega_\gamma) = 
   \frac{dq_\gamma}{d\epsilon_\gamma}(\theta,\phi)
   \left(\frac{dn_\gamma}{d\epsilon_\gamma}\right)^{-1}
   \times \left\{
\begin{array}{ll}
\cos^{-1}\theta,                     & \mbox{for the transparent disk};\\
1,                                   & \mbox{for the `isotropic' disk};\\
\cos\theta,                          & \mbox{for the emitting surface},
\end{array}
\right.
\end{equation}
where the function $dq_\gamma(\theta,\phi)/d\epsilon_\gamma$ is the
differential emissivity (stars and dust) of the disk (i.e., number of
photons emitted per unit time per unit area per unit solid angle per
unit energy interval) as seen from the point $(0,0,z)$.  For the case
of a homogeneous distribution $dq_\gamma/d\epsilon_\gamma = const$.  If
the distribution of the emissivity in the plane is symmetrical about
the disk axis then $dq_\gamma(\theta,\phi)/d\epsilon_\gamma \equiv
dq_\gamma(r[\theta,\phi])/d\epsilon_\gamma$ is a function of the
distance from the axis expressed in terms of the angles,
$r(\theta,\phi) = (\rho^2 +z^2\tan^2\theta +2\rho z \tan\theta
\cos\phi)^{1/2}$.

The differential number density of background photons at any point
$(0,0,z)$ is given by
\begin{equation}
\label{IC.23}
\frac{dn_\gamma}{d\epsilon_\gamma} =
   \int_{c_1}^{c_2} d\cos\theta \int_{\phi_1}^{\phi_2} d\phi\,
   \frac{dq_\gamma}{d\epsilon_\gamma}(\cos\theta,\phi) \times
\left\{
\begin{array}{c}
\cos^{-1}\theta\\ 1\\ \cos\theta
\end{array}
\right\}
,
\end{equation}
where $c_{1,2}, \phi_{1,2}$ are defined in
eqs.~(\ref{IC.19})--(\ref{IC.40}). For the simplest case
$dq_\gamma/d\epsilon_\gamma = C = const$, at a point slightly above the
disk center ($\phi_1 = 0, \phi_2 = 2\pi, c_1 \to +0, c_2 = 1$) one can
obtain
\begin{equation}
\label{IC.33}
\frac{dn_\gamma}{d\epsilon_\gamma} =
   C (\phi_2-\phi_1)\times 
\left.\left\{
\begin{array}{c}
\log s\\ s\\ s^2/2
\end{array}
\right\}\right|_{s=c_1}^{s=c_2} 
=
\left\{
\begin{array}{c}
-2 \pi C \log c_1 \\ 2 \pi C \\ \pi C
\end{array}
\right\} 
,
\end{equation}
where the infinity in the top row follows from using the thin disk
approximation and does not create any singularity in the formulas
since it appears only for points at the disk plane, $z=0$ (see also
Section \ref{discussion}).

\subsection{Scattering off a spectral distribution of electrons and 
source photons}

The final \gray emissivity spectrum (photons cm$^{-3}$ s$^{-1}$
sr$^{-1}$ MeV$^{-1}$) can be obtained after integration over the energy
spectra of electrons and the background radiation
\begin{equation}
\label{IC.30}
\frac{dF}{d\epsilon_2} =
   \frac{n_e n_\gamma c}{4\pi m_ec^2}
   \int d\epsilon_1 \int d\gamma \, \epsilon_1^2 \gamma^2
   f_\gamma(\epsilon_1) f_e(\gamma)
   \frac{dR(\gamma,\epsilon_1)}{d\epsilon_2},
\end{equation}
where $(n_\gamma \epsilon_1^2 f_\gamma)$, $(n_e
\gamma^2 f_e)$ are the differential number densities of photons and
electrons, respectively. The resulting spectrum depends strongly on
$\epsilon_2$, $f_{\gamma,e}$, and spatial coordinates.  However, we
find that for the power-law electron spectrum, at least, the ratio
\begin{equation}
\label{IC.31}
\Re = \frac{dF/d\epsilon_2}{dF_{iso}/d\epsilon_2}
\end{equation}
is much less sensitive and depends mostly on the spatial variables. The
value of $dF_{iso}/d\epsilon_2$ is calculated in the same way as
eq.~(\ref{IC.30}), but instead of $dR/d\epsilon_2$ the isotropic
function is used $dR_{iso}/d\epsilon_2$ (eq.~[\ref{IC.32}]).
Therefore, for particular geometry and photon and electron spectra,
the value $\Re$ gives essentially the geometrical factor.

\begin{figure}[t]
\centerline{
      \includegraphics[width=\fwb]{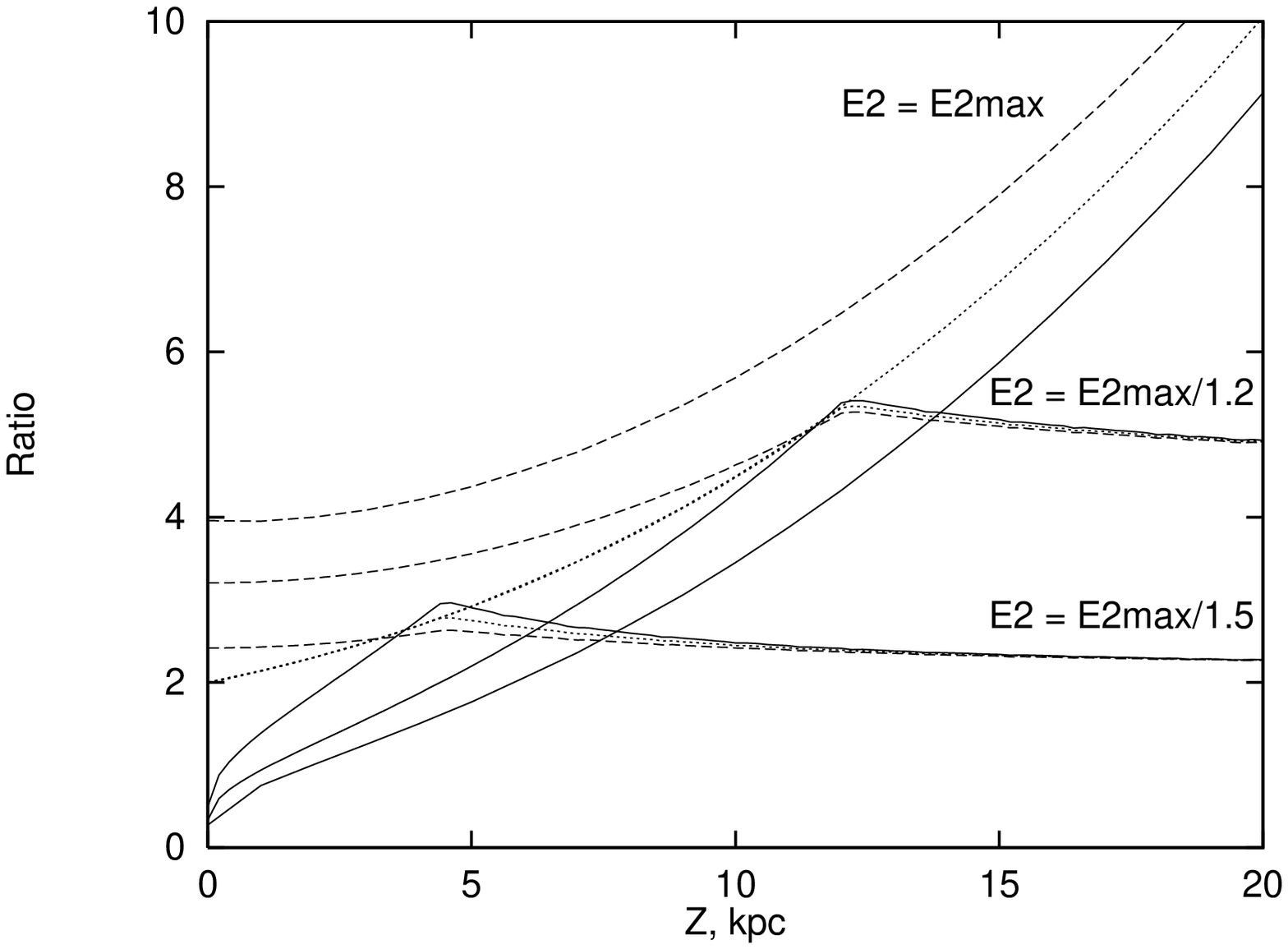}
      \hspace{\hs}
      \includegraphics[width=\fwb]{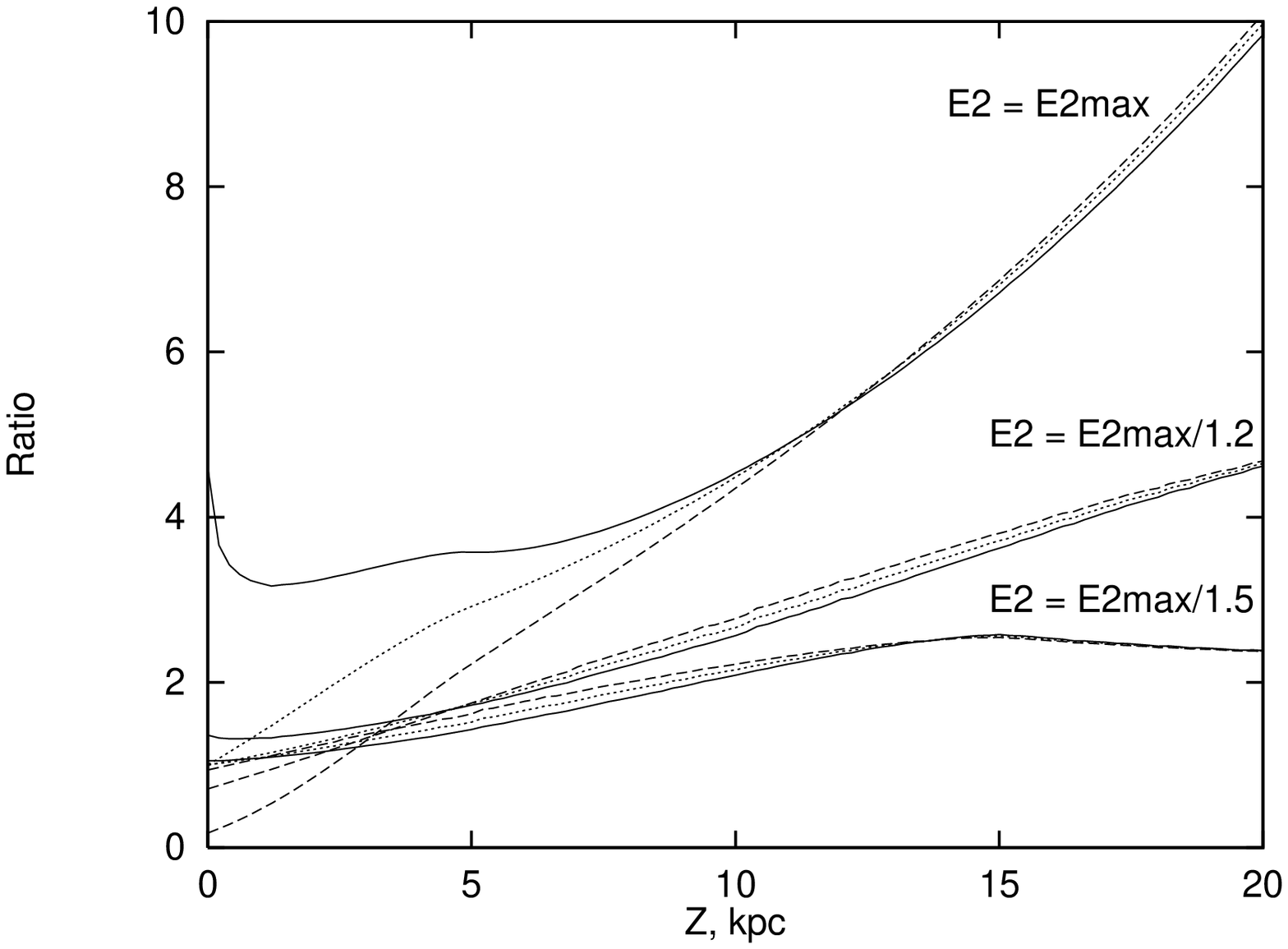}
}
\figcaption[fig3a.ps,fig3b.ps]{
The ratio $\frac{dR}{d\epsilon_2}/\frac{dR_{iso}}{d\epsilon_2}$
(eqs.~[\ref{IC.8}], [\ref{IC.32}]) vs.\ $z$ for a small volume of
isotropically distributed electrons at the disk axis for an observer at
the center of the Galaxy (left), and at the solar position $R_s = 8.5$
kpc (right).  Dotted lines show the ratio for the `isotropic' disk,
solid lines for the transparent disk (the Galaxy), and dashes for the
emitting surface case.  Other parameters:  $R_d = 15$ kpc, $\epsilon_1
= 2\times 10^{-7}$, $\gamma = 10^5$.
\label{fig3}}
\end{figure}

\placefigure{fig3}

\subsection{Calculations and Discussion} \label{discussion}
The effect described has a clear geometrical interpretation.  Jones'
(1968) formula assumes an isotropic distribution of photons while in our
case the photons are concentrated in a solid angle $\Omega_\gamma$
which is defined by the geometry.  Therefore, $4\pi/\Omega_\gamma$
gives a rough estimate of the enhancement when the effect of
anisotropy is taken into account. This is particularly true when one
considers the maximum energy of the upscattered photons $\epsilon_2=
\epsilon_{2\max}(\gamma,\epsilon_1)$ which can be reached for given
$\epsilon_1$ and $\gamma$ (eq.~[\ref{IC.41}]).  At energies
$\epsilon_2< \epsilon_{2\max}$, this effect is smaller since the
angular distribution of the background photons becomes less important
(larger scattering angles in the ERS become allowed). At very small
energies $\epsilon_2 \ll \epsilon_{2\max}$, the effect disappears
completely (anisotropic = isotropic) since this corresponds to the case
when the background photons are scattered at large angles, and their
actual angular distribution is of no importance.

Fig.~\ref{fig3} shows the spectral emissivity ratio
$\frac{dR}{d\epsilon_2}/\frac{dR_{iso}}{d\epsilon_2}$ of anisotropic to
isotropic scattering (eqs.~[\ref{IC.8}], [\ref{IC.32}]), vs.\ $z$ for
electrons at the disk axis ($\rho = 0$) for an observer in the center
of the emitting disk (left), and at the solar position $R_s = 8.5$ kpc
(right).  We show three cases:  transparent disk (the Galaxy), emitting
surface, and `isotropic' disk.  The other parameters are:  $R_d = 15$
kpc, $\epsilon_1 = 2\times 10^{-7}$, $\gamma = 10^5$.  To illustrate
the energy dependence of the ratio, the curves are shown for three
different photon energies:  $\epsilon_2\approx \epsilon_{2\max}$,
$\epsilon_{2\max}/1.2$, and $\epsilon_{2\max}/1.5$.  In the simplest
geometry (left panel) when the scattering electrons and the observer
are both on the disk axis, the `isotropic' disk case gives exactly the
same enhancement as obtained from the rough estimate
$4\pi/\Omega_\gamma$. This reflects simply the larger number of
background photons (by a factor of $4\pi/\Omega_\gamma$) per solid
angle compared to the isotropic case.  Other cases, the transparent
disk and emitting surface, give somewhat different results because of
the more complicated angular distributions of the background photons.
The ratio increases with $z$ until the whole disk is covered by the
solid angle $\Omega_\gamma$, which is defined by
eqs.~(\ref{IC.14})--(\ref{IC.16}), and then remains almost a constant.

The importance of the angular distribution of the background photons is
clear from comparison of the behaviour of the ratio at small $z$ on the
left and right panels (Fig.~\ref{fig3}).  In the case of the
transparent disk, the photon distribution is $\sim \cos^{-1}\theta$
which has a minimum at $\theta=0$.  The solid line in the left panel
(scattering electrons and the observer are both on the disk axis) thus
shows the smallest effect.  The picture changes when the scattering
electrons do not lie just above the observer (right panel).  In this
case, $z \to 0$ corresponds to $\theta \to \pi/2$ where the angular
distribution has an infinite peak.  Therefore, in the case
$\epsilon_2\approx \epsilon_{2\max}$, the ratio $\to \infty$ as $z \to
0$. This infinity, however, is not physical since it is connected with
approximation of infinitely thin disk plane.  The result can be
compared with the emitting surface case ($\sim \cos\theta$; dashed
line) where the opposite, but finite, effect is seen.

\begin{figure}[t]
\centerline{
      \includegraphics[width=\fwb]{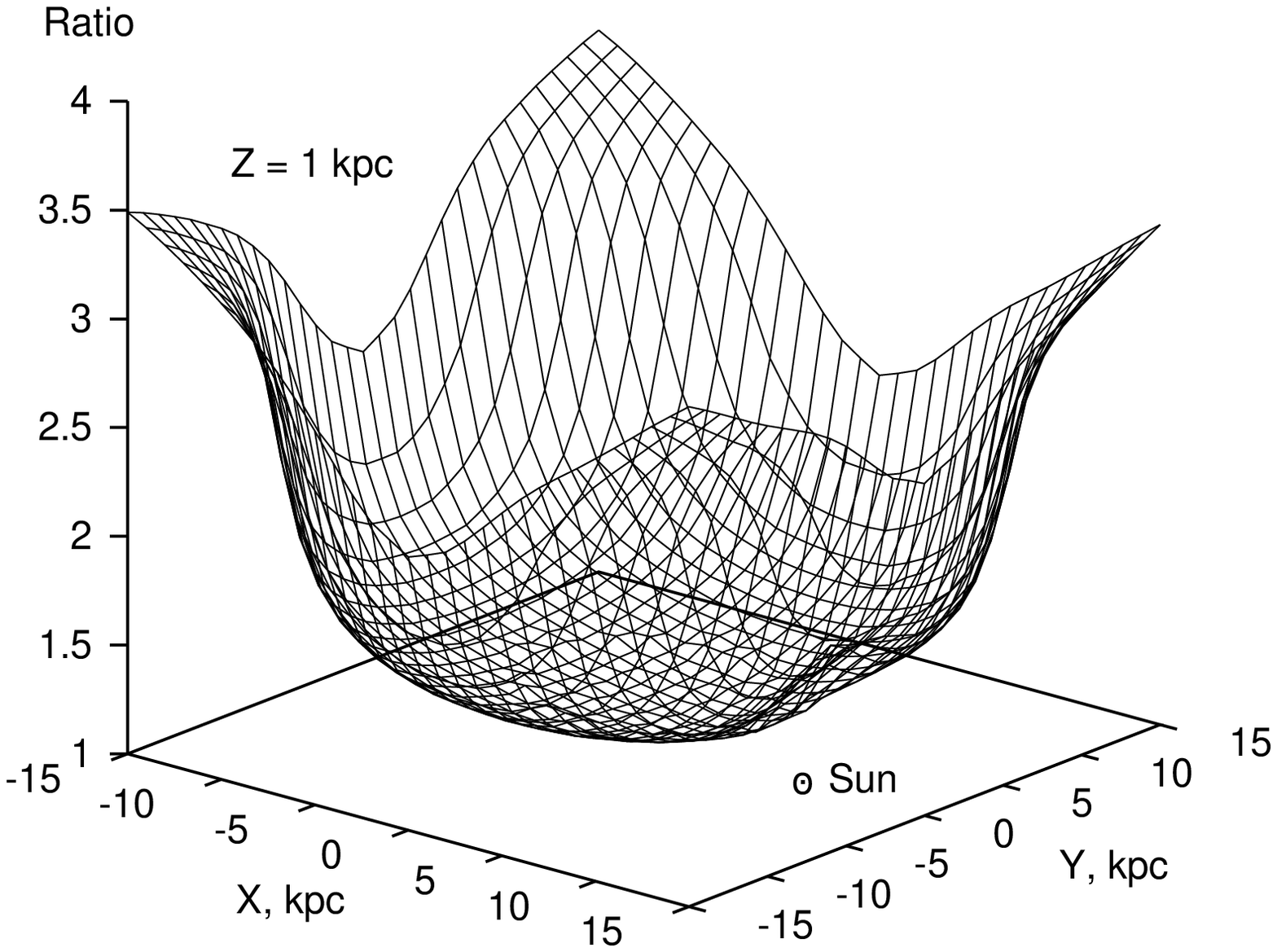}
      \hspace{\hs}
      \includegraphics[width=\fwb]{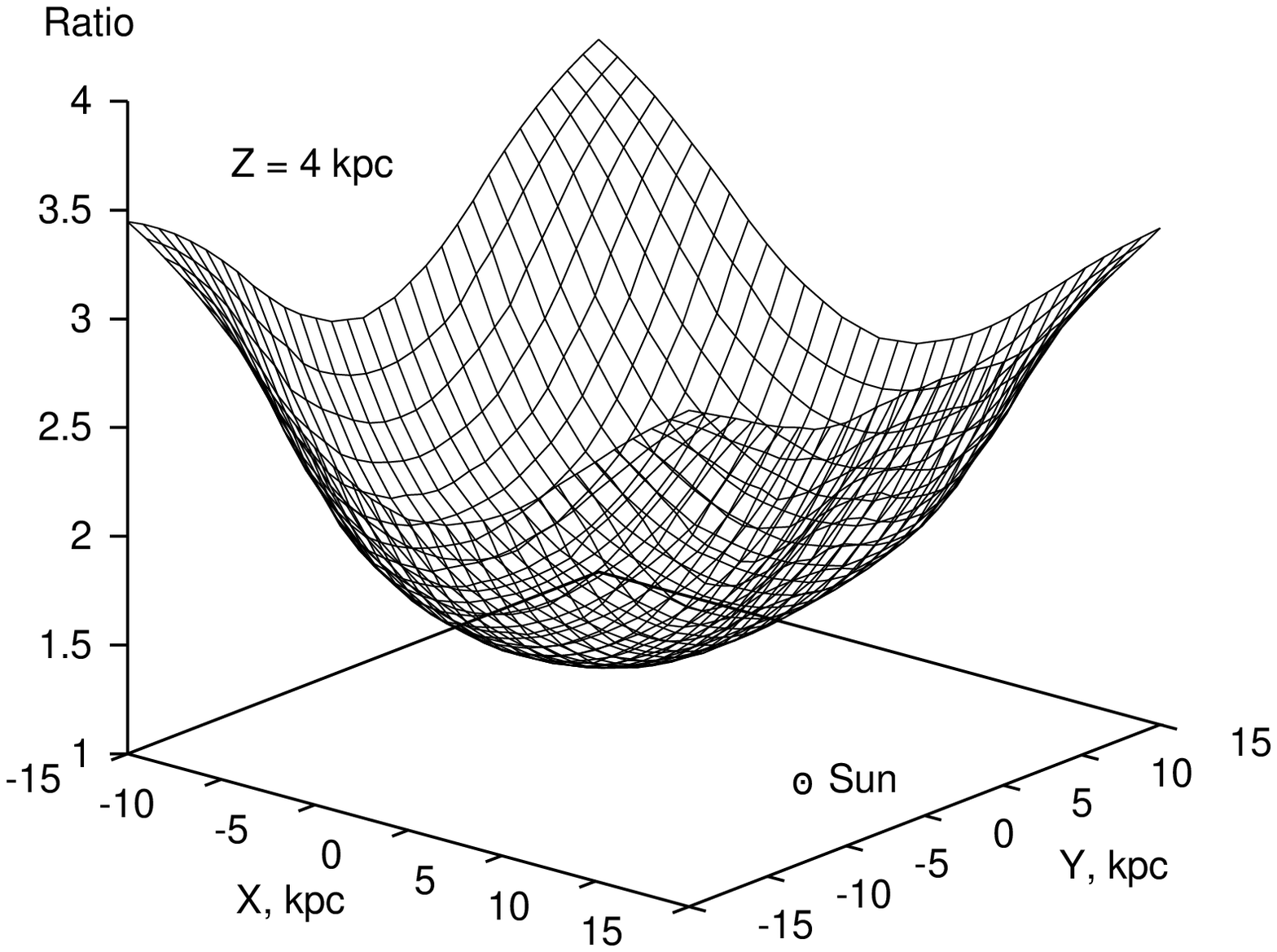}
}
\centerline{
      \includegraphics[width=\fwb]{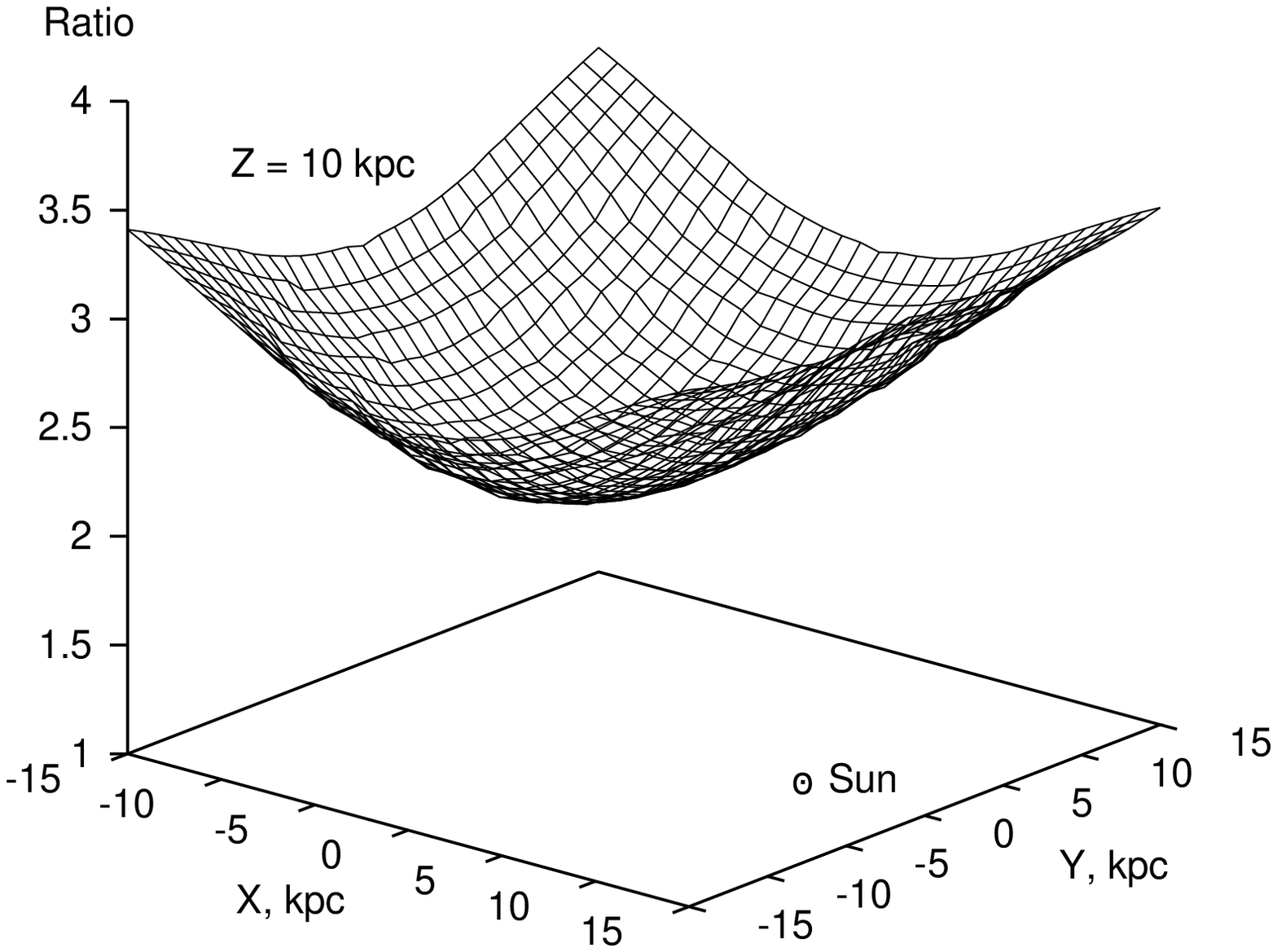}
      \hspace{\hs}
      \includegraphics[width=\fwb]{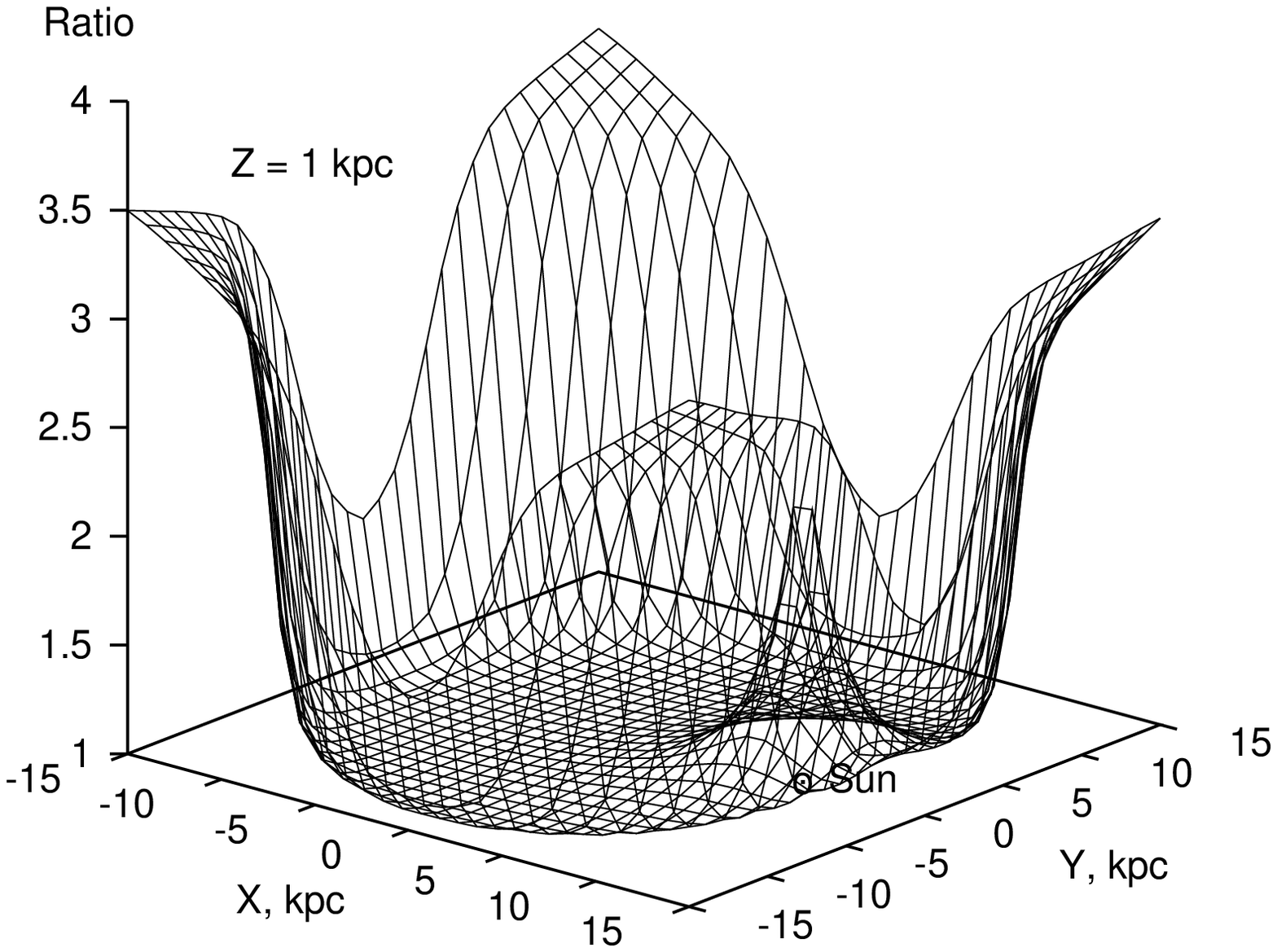}
}
\figcaption[fig4a.ps,fig4b.ps,fig4c.ps,fig4d.ps]{
Anisotropic/isotropic ICS ratio $\Re$ (eq.~[\ref{IC.31}]) as calculated
for an observer at $R_s = 8.5$ kpc from Galactic center. Other
parameters are: $R_d = 15$ kpc, electron power-law index $-3$,
$\epsilon_1 = 1$ eV and $\epsilon_2 = 100$ MeV, for $z = 1$ kpc (top
left), 4 kpc (top right), and 10 kpc (bottom left).  The bottom right
panel shows a calculation for the emitting surface case ($z = 1$ kpc).
\label{fig4}}
\end{figure}

\placefigure{fig4}

Fig.~\ref{fig4} shows $\Re$ (eq.~[\ref{IC.31}]) as calculated for a
homogeneous transparent disk with radius $R_d = 15$ kpc, $\epsilon_2 =
100$ MeV, monoenergetic background photons $\epsilon_1 = 1$ eV, an
electron spectral power-law index of $-3$, and for $z = 1$, 4, and 10
kpc.  At small distances from the plane within the disk the ratio is
close to unity since the emitting electrons directed to the observer
are moving under small angles to the disk plane and the radiation field
they see is effectively isotropic.  As the ICS electrons become higher
in altitude and/or further from the disk center the radiation field
they see becomes more and more anisotropic, and therefore the ratio
$\Re$ increases.

The smaller value of the effect compared to that shown in
Fig.~\ref{fig3} is the result of integration over the electron
spectrum.  As mentioned above the effect is maximal when
$\epsilon_2= \epsilon_{2\max}$ or equivalently $\gamma=
\gamma_{\min}(\epsilon_1,\epsilon_2)= \frac{1}{2}(\epsilon_2
+[\epsilon_2^2 +\epsilon_2/\epsilon_1]^{1/2})$. At $\gamma >
\gamma_{\min}$ the effect is smaller, but the background photons come
from a larger solid angle to be upscattered to the same energy; this
increases the number of background photons involved.  The actual value
of the enhancement is then naturally obtained from a balance between
the decreasing number of electrons at higher energies and the
increasing value of the effective solid angle allowed for the
background photons.

Fig.~\ref{fig4} (bottom right) shows as an example a calculation made
for the emitting surface ($z = 1$ kpc).  An enhancement of the emission
above the solar position is clearly seen. 

\pagebreak[3]
\section{Diffuse Galactic $\gamma$-rays}
In the following the reader is referred to \cite{SMR99} for details of
the models.

The interstellar radiation field (ISRF) is essential for propagation
and \gray production by CR electrons. It is made up of
contributions from starlight, emission from dust, and the cosmic
microwave background. Estimates of the spectral and spatial
distribution of the ISRF rely on models of the distribution of stars,
absorption, dust emission spectra and emissivities, and is therefore in
itself a complex subject.

Recent data from infrared surveys by the IRAS and COBE satellites have
greatly improved our knowledge of both the stellar distribution and the
dust emission.  A new estimate of the ISRF has been made based on these
surveys and stellar population models (\cite{SMR99}).  Stellar emission
dominates from 0.1 $\mu$m to 10 $\mu$m, emission from very small dust
grains contributes from 10 $\mu$m to 30 $\mu$m.  Emission from dust at
$T \sim20$ K dominates from 30 $\mu$m to 400 $\mu$m.  The 2.7 K
microwave background is the main radiation field above 1000 $\mu$m.

The ISRF has a vertical extent of several kpc since the Galaxy acts as
a disk-like source of radius $\sim$15 kpc. The radial distribution of
the stellar component is centrally peaked because the stellar
density increases exponentially inwards with a scale-length of
$\sim$2.5 kpc until the bar is reached. The dust component is related
to that of the neutral gas (HI + H$_2$) and is therefore distributed
more uniformly in radius than the stellar component.

\begin{figure}[t]
\centerline{
      \includegraphics[width=\fwd]{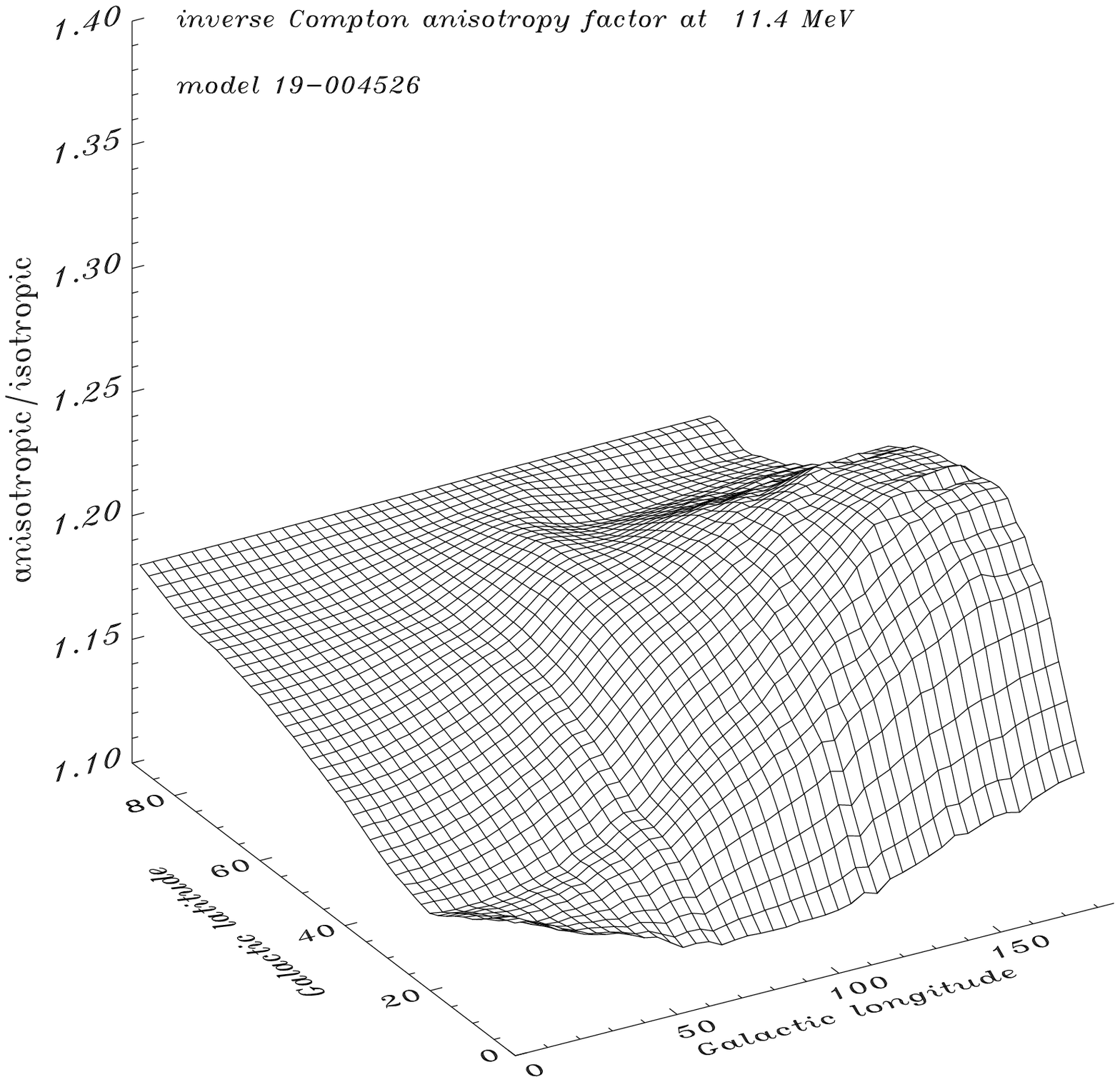}
      \hspace{\hs}
      \includegraphics[width=\fwd]{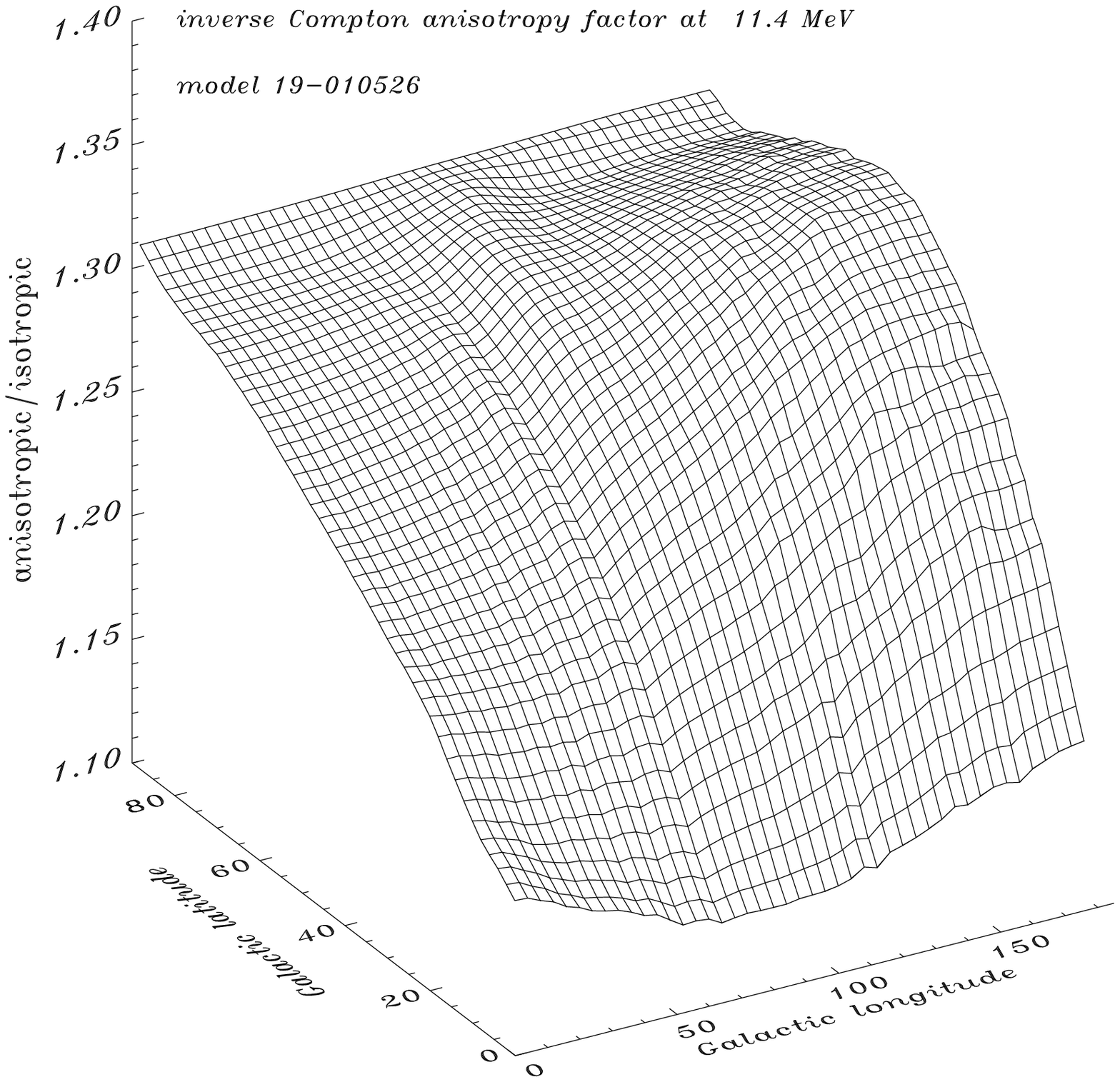}
}
\figcaption[fig5a.ps,fig5b.ps]{
Latitude -- longitude plot of the anisotropic/isotropic intensity ratio
for 11.4 MeV $\gamma$-rays (ratio of the two sky maps).  Halo size $z_h=4$
kpc (left) and 10 kpc (right).
\label{fig5}}
\end{figure}

\placefigure{fig5}

\begin{figure}[!t]
\centerline{
      \includegraphics[width=\fwb]{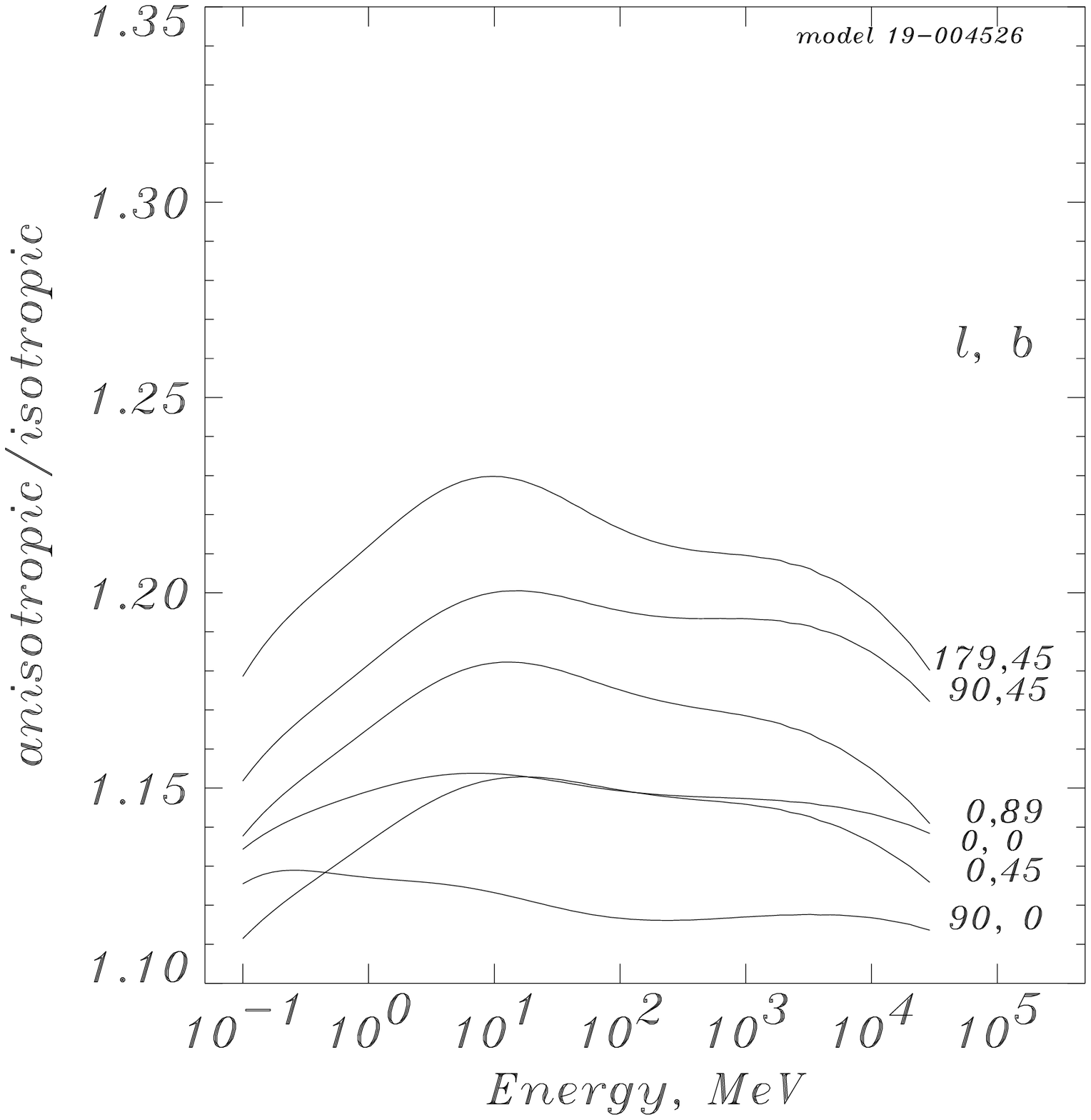}
      \hspace{\hs}
      \includegraphics[width=\fwb]{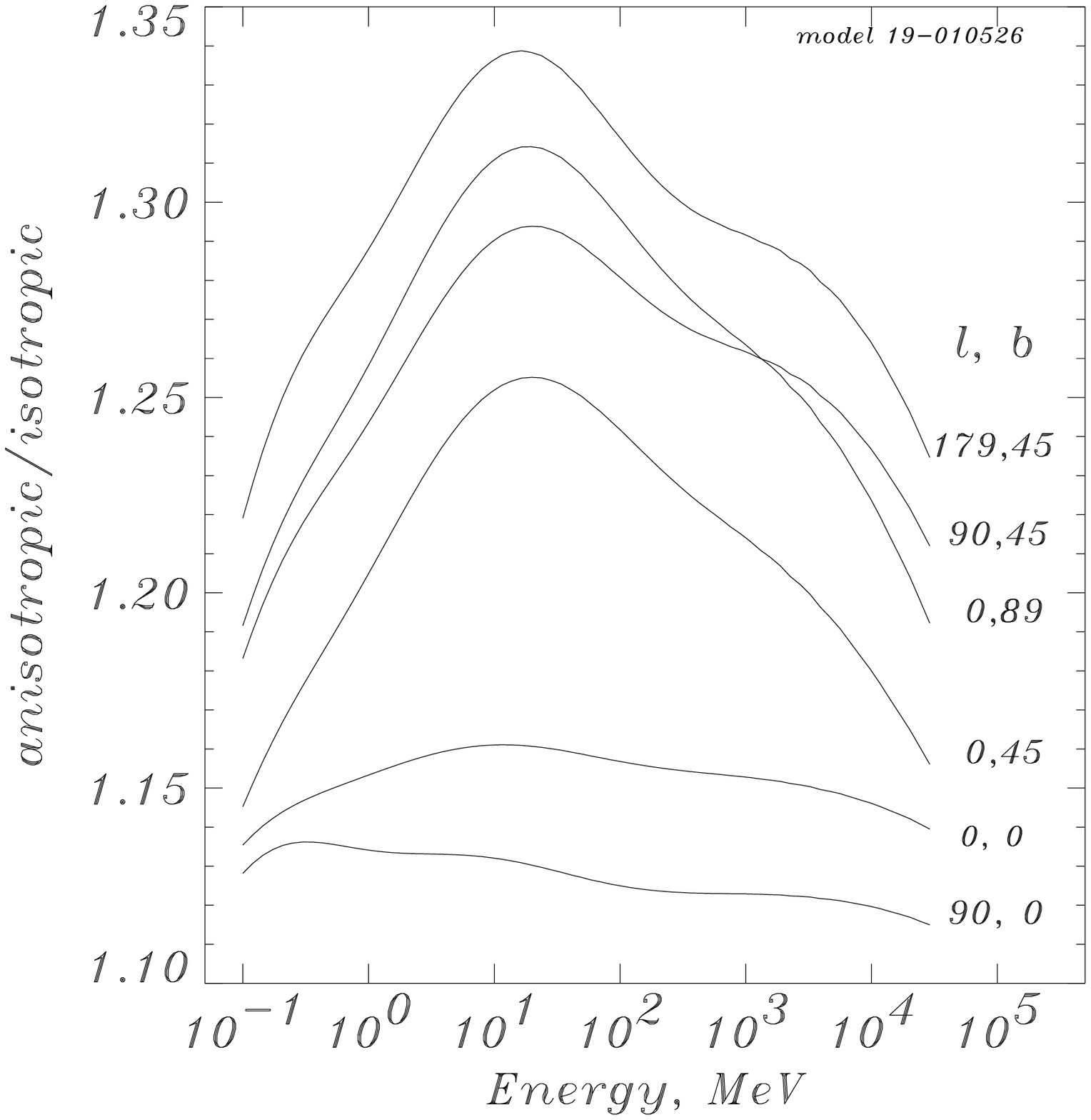}
}
\figcaption[fig6a.ps,fig6b.ps]{
The intensity ratio vs.\ \gray energy for some direction as seen from
the solar position. The corresponding Galactic coordinates $(l,b)$ are
shown near the right scale.  Halo size $z_h=4$ kpc (left) and 10 kpc
(right).
\label{fig6}}
\end{figure}

\placefigure{fig6}

In practice we calculate the anisotropic/isotropic ratio $\Re$ for any
particular model of the particle propagation (halo size, electron
spectral injection index etc.) on a spatial grid taking into account
the difference between stellar and dust contributions to the ISRF, and
then interpolate it when integrating over the line of sight 
(see \cite{SMR99}).

Fig.~\ref{fig5} shows a Galactic latitude -- longitude plot of the
intensity ratio for 11.4 MeV \grays for two Galactic models with halo
size $z_h=4$ kpc and 10 kpc.  This is obtained from the computed sky
maps in the anisotropic and isotropic cases. The calculation has been
made with a `hard' interstellar electron spectrum (the interstellar
electron spectrum is discussed below).  It is seen that the enhancement
due to the anisotropic ICS can be as high as a factor $\sim$1.4 for the
pole direction in models with a large halo, $z_h\ga 10$ kpc.  The
maximal enhacement occurs at intermediate Galactic latitudes in the
outer Galaxy because there head-on collisions dominate.
Fig.~\ref{fig6} shows the intensity ratio vs.\ \gray energy for several
directions as seen from the solar position for a halo size $z_h=4$ kpc
and 10 kpc.

\begin{figure}[!t]
\centerline{
      \includegraphics[width=\fwb]{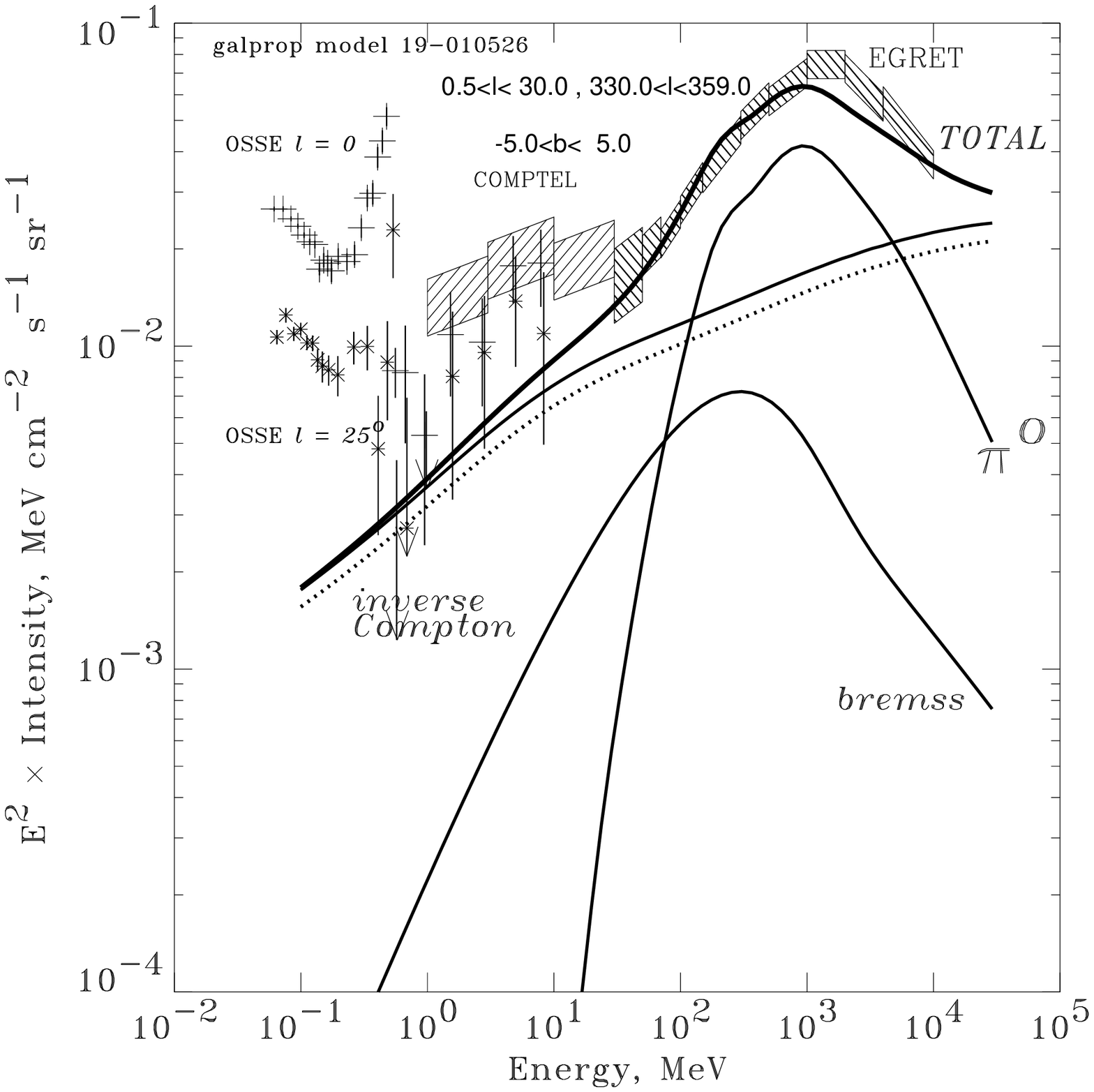}
      \hspace{\hs}
      \includegraphics[width=\fwb]{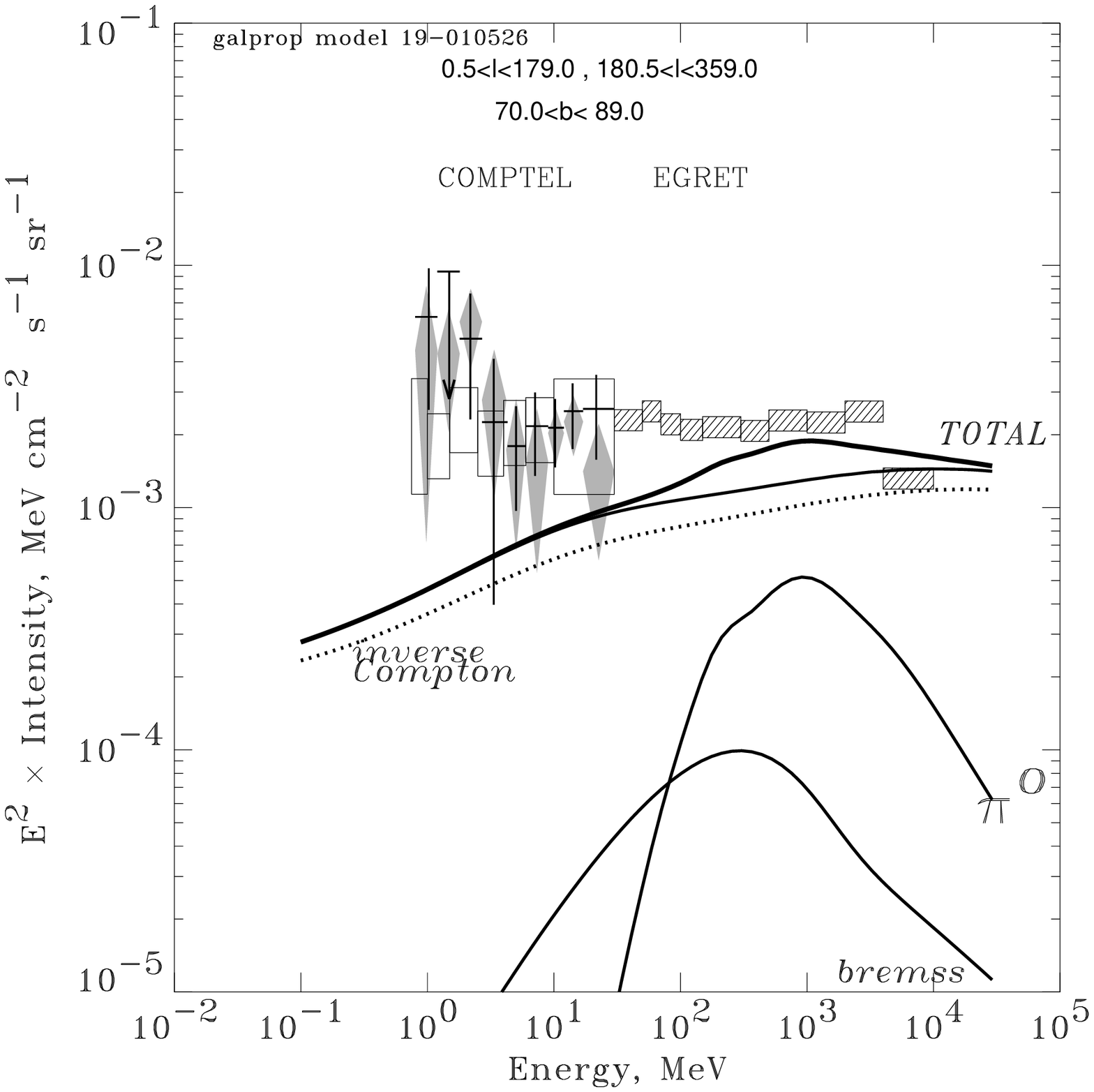}
}
\figcaption[fig7a.ps,fig7b.ps]{
Gamma-ray energy spectrum of the inner Galaxy (left panel) and from
high Galactic latitudes ($|b|\ge 70^\circ$, all longitudes) (right
panel) as compared with calculations in the HELH model.  Separate
components show the contribution of bremsstrahlung, $\pi^0$-decay, and
ICS with (solid line) and without (dotted line) the anisotropic
effect.  Data in the left panel: EGRET (\cite{StrongMattox96}), 
COMPTEL (\cite{Strongetal99}), and OSSE (\cite{Kinzer99}). Data in the
right panel: COMPTEL high-latitude total intensity (open boxes:
\cite{Bloemen99}, diamonds:  \cite{Kappadath98}, crosses:
\cite{Weidenspointner99}), and total from EGRET measurements
(\cite{SMR99}).
\label{fig7}}
\end{figure}

\placefigure{fig7}

\begin{figure}[t]
\centerline{
      \includegraphics[width=\fwb]{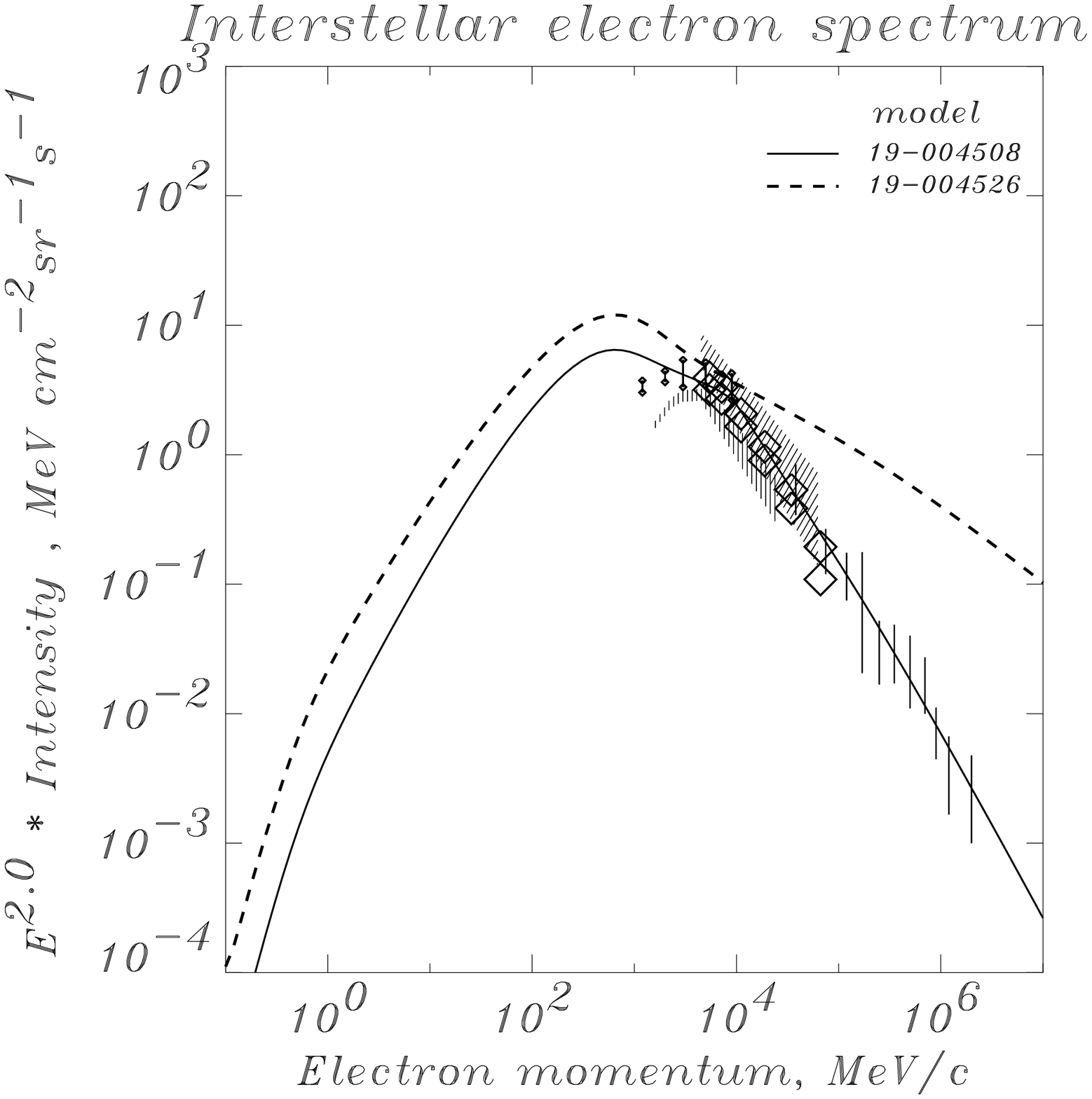}
      \hspace{\hs}
      \includegraphics[width=\fwb]{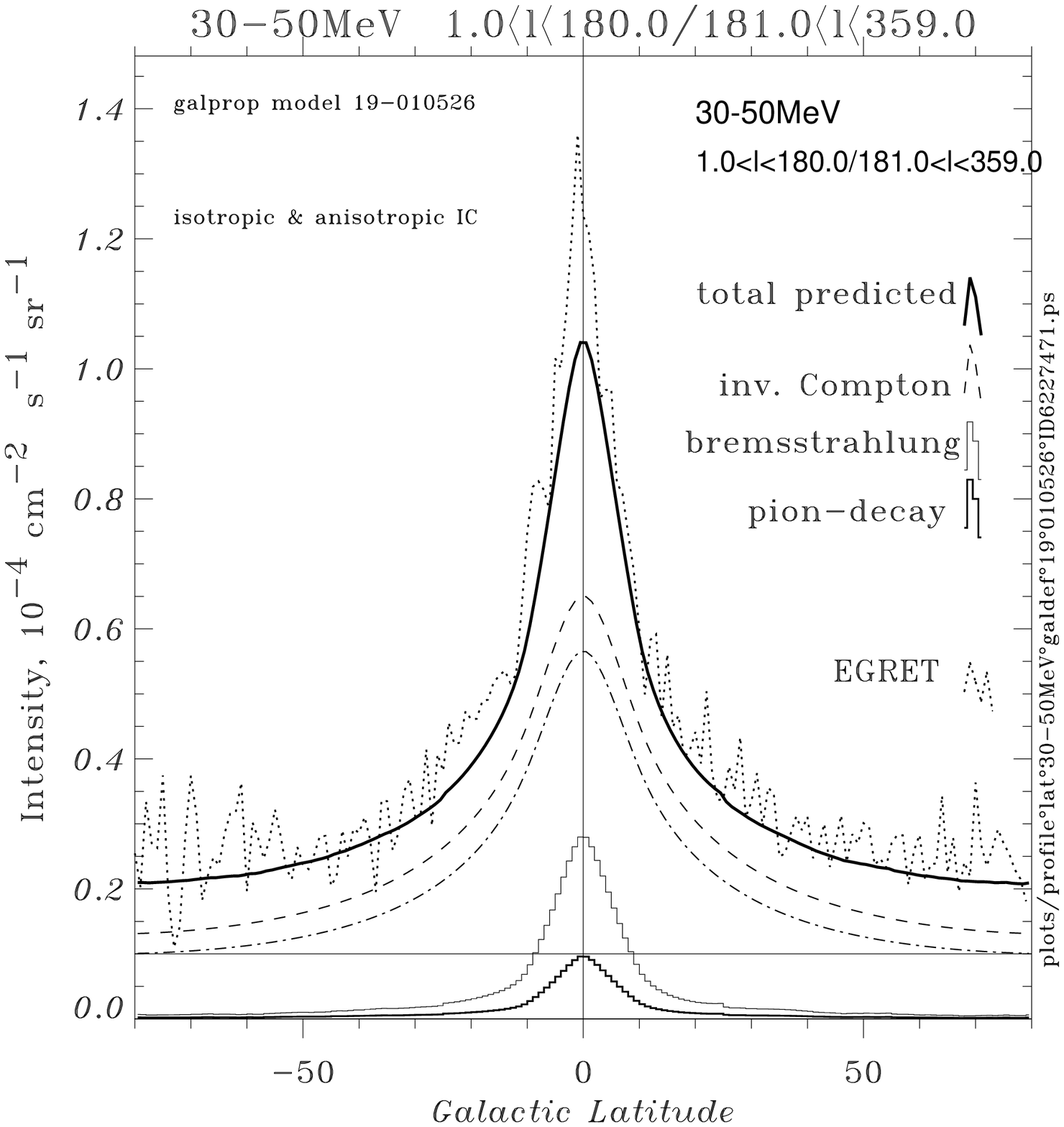}
}
\parbox{89mm}{%
\figcaption[fig8.ps]{
Electron spectra at $R_s = 8.5$ kpc in the plane, for
`conventional' (solid line) and hard electron spectrum (dashes)
models.  Data points: direct measurements, see references in
\cite{MS98a}. 
\vspace{3\baselineskip}\vspace{1pt}
\label{fig8}}
}\hspace{7mm}
\parbox{89mm}{%
\figcaption[fig9.ps]{
The latitude 30--50 MeV \gray profile of the Galaxy as calculated in
the HELH model (heavy line; convolved) compared to EGRET data (dots;
point sources removed).  Separate components show the contribution of
bremsstrahlung (thin histogram), $\pi^0$-decay (thick histogram), and
ICS with (dashes) and without (dash-dots) anisotropic effect;
horizontal line: isotropic background.
\label{fig9}}
}
\end{figure}

\placefigure{fig8}
\placefigure{fig9}

Fig.~\ref{fig7} shows the spectra of the inner Galaxy ($|b| \le
5^\circ$, $330^\circ\le l\le 30^\circ$) and for high Galactic latitudes
($|b|\ge 70^\circ$, all longitudes) as calculated for the HELH model.
The HELH model (`hard electron spectrum and a large halo',
\cite{SMR99}) is chosen to fit high-energy \grays using hard electron
spectrum\footnote{  In fact, the size of the anisotropic effect
will increase for a softer electron spectrum (see
Section~\ref{discussion}).   } (the injection spectral index is taken
as --1.8, with reacceleration) and a broken nucleon spectrum (injection
spectral index is --1.8/--2.5 with break at 20 GeV/nucleon, with
reacceleration).  Such a nucleon spectrum satisfies the limits imposed
by antiprotons and positrons (\cite{MSR98}, \cite{SMR99}).  The halo
size taken, $z_h=10$ kpc, is within the limits (4--12 kpc) obtained
from our $^{10}$Be studies (\cite{SM98}).  Following Pohl \& Esposito
(1998), the consistency with the locally measured electron spectrum
above 10 GeV is {\it not} required, since the rapid energy losses cause
a clumpy distribution so that this is not necessarily representative of
the interstellar average.  For this case, the interstellar electron
spectrum deviates strongly from that locally measured as illustrated in
Fig.~\ref{fig8}.  Because of the increased ICS contribution at high
energies, the predicted \gray spectrum can reproduce the overall
intensity of the inner Galaxy from 30 MeV to 10 GeV.
If the hypothesis of a hard electron injection spectrum is correct in
explaining the large-scale Galactic emission, then the same model for
the electrons is applicable to the halo where large-scale propagation is
relevant. We consider a wide high-latitude region $70^\circ<b<90^\circ$
which encloses a large volume of halo emission and is therefore unaffected
by short-scale fluctuations.

For comparison the dotted lines in Fig.~\ref{fig7} show the spectra of
ICS \grays as calculated using the isotropic Jones' formula
(eq.~[\ref{IC.32}]).  The model with anisotropic ICS component gives
somewhat larger fluxes.  The effect is larger at high Galactic
latitudes, but also significant for the inner Galaxy.  For the case of
the inner Galaxy the result is not sensitive to the halo size and for
smaller halos the plot would be almost the same.  At high latitudes,
where the ICS is the only important contributor into the Galactic
diffuse emission, it provides up to a factor $\sim1.4$ larger flux
compared to isotropic calculations.

Fig.~\ref{fig9} shows the \gray latitude profile of the Galaxy for
30--50 MeV as calculated in the HELH model, again with and without the
anisotropic effect, compared to the EGRET data.  For this comparison
the \gray sky maps computed in our model have been convolved with the
EGRET point-spread function; point sources contribution has been
removed from the EGRET data (see \cite{SMR99} for details).  At high
latitudes the model with anisotropic ICS component again provides
significantly larger fluxes compared to isotropic calculations.  This
is particularly important for estimates of the extragalactic background
since Galactic ICS and extragalactic contributions may be comparable.

\section{Conclusion}
The ICS is a major contributor in the diffuse Galactic emission above
$\sim 1$ MeV.  Since accurate observational data from EGRET is now
available it is desirable to compute the ICS including all  important
effects.  We have shown that the anisotropy of the interstellar
radiation field has a significant effect on the intensity and angular
distribution of \gray radiation, and can increase the high-latitude
Galactic \gray flux up to 40\%.  This effect should be taken into
account when calculating the Galactic emission for extragalactic
background estimates.

The derivations presented here provide the basis for the treatment of
ICS in our companion study of Galactic diffuse continuum emission
(\cite{SMR99}).

\twocolumn
\nopagebreak



\end{document}